\documentclass[twocolumn,prl,preprintnumbers,amsmath,amssymb,superscriptaddress]{revtex4}

\usepackage{graphicx}
\usepackage{dcolumn}
\usepackage{bm}

\usepackage[colorlinks=true,linktocpage=true,linkcolor=blue,citecolor=blue]{hyperref}

\graphicspath{{figures/}}

\begin{document}

\title{Mesonic Condensation in Isospin Matter under Rotation}

\author{Hui Zhang}
\affiliation{Institute of Particle Physics (IOPP) and Key Laboratory of Quark and Lepton Physics (MOE),  Central China Normal University, Wuhan 430079, China}
\address{Physics Department and Center for Exploration of Energy and Matter, Indiana University, 2401 N Milo B. Sampson Lane, Bloomington, IN 47408, USA.}

\author{Defu Hou}
\email{houdf@mail.ccnu.edu.cn}
\affiliation{Institute of Particle Physics (IOPP) and Key Laboratory of Quark and Lepton Physics (MOE),  Central China Normal University, Wuhan 430079, China}

\author {Jinfeng Liao} \email{liaoji@indiana.edu}
\address{Physics Department and Center for Exploration of Energy and Matter, Indiana University, 2401 N Milo B. Sampson Lane, Bloomington, IN 47408, USA.}
\affiliation{Institute of Particle Physics (IOPP) and Key Laboratory of Quark and Lepton Physics (MOE),  Central China Normal University, Wuhan 430079, China}

\date{\today}

\begin{abstract}
We  investigate the mesonic condensation in isospin matter under rotation. Using the two-flavor NJL effective model under the presence of global rotation, we  demonstrate two important effects of the rotation on its phase structure: a rotational suppression of the scalar-channel condensates, in particular the pion condensation region; and a rotational enhancement of the rho condensation region with vector-channel condensate. A new phase diagram for isospin matter under rotation is   mapped out on the $\omega-\mu_I$ plane  where three distinctive phases, corresponding to $\sigma$, $\pi$, $\rho$ dominated regions respectively, are separated by a second-order line at low isospin chemical potential and a first-order line at high rotation which are further connected at  a tri-critical point.   \end{abstract}
\keywords{Isospin Matter, Rotation, QCD Phase Diagram}
\maketitle

\section{\label{sec:Intro} Introduction}

Recently there have been rapidly growing interests in understanding the properties and phase structures of matter under extreme fields like magnetic field or global rotation~\cite{Miransky:2015ava,Fukushima:2018grm}. Examples of such physical systems come from a variety of different areas, such as the hot quark-gluon plasma in peripheral heavy ion collisions~\cite{STAR:2017ckg,Becattini:2016gvu,Csernai:2013bqa,Becattini:2015ska,Jiang:2016woz,Shi:2017wpk,Deng:2016gyh,Pang:2016igs,Xia:2018tes}, dense nuclear matter in rapidly spinning neutron stars~\cite{Watts:2016uzu,Grenier:2015pya,Berti}, lattice gauge theory in rotating frame~~\cite{Yamamoto:2013zwa}, cold atomic gases~\cite{Fetter:2009zz,2008PhRvA..78a1601U,2009PhRvA..79e3621I} as well as certain condensed matter materials~\cite{spinhydro,Gooth:2017mbd}.

Rotation provides an interesting new type of macroscopic control parameter, in addition to conventional ones such as temperature and density. It has nontrivial interplay with microscopic spin degrees of freedom through the rotational polarization and could induce novel phenomena. For example, there are  anomalous transport effects such as the chiral vortical effect and chiral vortical wave in rotating fluid with chiral fermions~\cite{Son:2009tf,Kharzeev:2010gr,Landsteiner:2011iq,Hou:2012xg,Jiang:2015cva,Flachi:2017vlp,Kharzeev:2015znc}. Furthermore, if the underlying materials contain fermions that may form condensate via pairing, their phase structure can be significantly influenced by  global rotation~\cite{Jiang:2016wvv,Ebihara:2016fwa,Chen:2015hfc,Mameda:2015ria,Huang:2017pqe,Liu:2017spl,Liu:2017zhl,Chernodub:2016kxh,Chernodub:2017ref,Chernodub:2017mvp,Zubkov:2018gmc,Wang:2018sur,Wang:2018zrn,Wang:2019nhd}. A generic effect is the rotational suppression of  fermion pairing in the zero angular momentum states, as demonstrated for e.g. chiral phase transition and color superconductivity in strong interaction  systems~\cite{Jiang:2016wvv}.
Given the suppression of scalar pairing states, it is natural to wonder what may happen to  pairing states with nonzero angular momentum e.g. spin-1 condensate of fermionic pairs. One would expect them to be enhanced by rotation which prefers states with finite angular momentum and tends to polarize it along the rotational axis. It is of great interest to examine this  in concrete physical systems. We note a relevant and novel phenomenon of charged vector meson condensation induced by strong magnetic fields
~\cite{Chernodub:2011mc,Chernodub:2012tf}.

In this Letter, we perform the first analysis for the influence of rotation on the phase structure of isospin matter --- Quantum Chromodynamics (QCD) matter at finite isospin density (or equivalently chemical potential) which implies an imbalance  between the u-flavor and d-flavor of quarks~\cite{Son:2000xc,Son:2000by}. Such isospin matter is relevant for understanding the properties of neutron star materials with a significant mismatch between the number of neutrons and protons and thus also between u-quarks and d-quarks. Also the dense matter created in low energy heavy ion collisions bears significant isospin density arising from stopping of initial beam nuclei. It is also possible to simulate isospin matter with two-component cold fermionic gases. This paper focuses on a theoretical understanding with a brief outline of potential experimental application which will be the topic of future study.  

One interesting phenomenon in isospin matter is pion condensation~\cite{He:2005nk,He:2005sp,Kang:2013bea,Mao:2012fx,Cao:2015xja,Brauner:2016lkh}: a chiral $\sigma$ condensate (from quark-anti-quark pairing in the scalar channel) at low isospin density changes into a pion condensate (from quark-anti-quark pairing in the pseudo-scalar channel) at high isospin density.  Since both are pairing states zero angular momentum, one would expect a rotational suppression effect on both. Furthermore, rotation may induce condensation in other mesonic channels arising from quark-anti-quark pairing in non-zero angular momentum states such as the $\rho$-channel.  We will perform the first systematic study of these possible mesonic condensations  simultaneously in the isospin matter under global rotation.  We will show that there are indeed suppression of scalar pairing and enhancement of vector pairing due to rotation, with the emergence of rho condensation phase. Such analysis will allow us to  map out a new phase diagram on the  rotation-isospin   parameter plane with highly nontrivial phase structures. 
Note the system considered here consists of rotating normal components and irrotational condensates. The formation of condensate vortices is an interesting possibility for further study.

\section{\label{sec:Form} Formalism}

To investigate the mesonic condensation in isospin matter, we will adopt a widely-used effective model, namely the two-flavor  Nambu-Jona-Lasinio (NJL) model with four-fermion interactions in various channels at finite isospin chemical potential $\mu_I$:
\begin{eqnarray} \label{eq_1}
{\cal L} &=& \bar\psi(i\gamma_\mu\partial^\mu-m_0+{\mu_I\over2}\gamma_0\tau_3)\psi
+ {\cal L}_{I}^{s} + {\cal L}_{I}^{v} \,\, , \\
{\cal L}_{I}^{s} &=& {G_s}\left[\left(\bar\psi\psi\right)^2+\left(\bar\psi i\gamma_5{\boldsymbol \tau}\psi\right)^2\right] \,\, , \\
{\cal L}_{I}^{v} &=& -{G_v} \left(\bar\psi \gamma_\mu{\boldsymbol \tau}\psi\right)^2  \,\, .
\end{eqnarray}
In the above, the $m_0= 5\rm MeV$ is the light quark mass parameter while $G_s=G_v=5.03\ {\rm GeV^{-2}}$ are the scalar and vector channel coupling constants respectively. The NJL-type effective model also requires a momentum cut-off parameter  $\Lambda=650\ {\rm MeV}$. These are standard choices, yielding a vacuum expectation value (VEV)  $\sigma_0=2\times(250\rm MeV)^3$. The NJL model  is not the same as QCD itself but nevertheless effectively captures its  low energy chiral dynamics.

We consider three possible mesonic condensation scenarios: condensation of $\sigma$, $\pi$ or $\rho$ fields respectively.  Following the standard mean-field method, we introduce:
\begin{eqnarray}
\sigma= \langle \bar\psi\psi \rangle,\
\pi= \langle \bar\psi i \gamma_5 {\tau_3}  \psi \rangle,\
\rho= \langle \bar\psi i \gamma_0 {\tau_3} \psi \rangle.
\end{eqnarray}
Note a possible extension of the present analysis is to include an axial vector channel coupling term into the Lagrangian Eq.(\ref{eq_1}). This may allow a new axial vector mean field condensate in competition with the above ones --- an interesting possibility for further study.   

Furthermore we are considering such a system under global rotation around $\hat{z}$-axis with angular velocity $\vec{\omega}=\omega \hat{z}$. To do this, one can study the system in the rotating frame and rewrite the spinor theory with the curved metric associated with the rotating frame~\cite{Jiang:2016wvv}. The main new effect is a global polarization term in the Lagrangian density:
\begin{eqnarray}
{\cal L}_{R} = \psi^\dagger \left[  (\vec{\omega}\times \vec{x})\cdot (-i \vec{\partial})
+ \vec{\omega} \cdot \vec{S}_{4\times 4} \right] \psi
\end{eqnarray}
where $\vec{S}_{4\times 4} = \frac{1}{2} Diag\left(\vec{\sigma} , \vec{\sigma} \right)$ is the spin operator with $\vec{\sigma}$ the $2\times 2$ Pauli matrices. 

Within mean-field approximation and assuming $\omega\ r \ll 1$, one obtains the following thermodynamic potential of NJL model for isospin matter under rotation:
\begin{eqnarray} \label{eq_Omega}
\Omega &=& G_s (\sigma^2+\pi ^2) -G_v \rho^2 \nonumber
\\ && -\frac{\rm{N_c N_f}}{16 \pi ^2} \sum _n \int {\rm d} k_t^2 \int {\rm d} k_z [J_{n+1}(k_t r)^2+J_n(k_t r)^2] \nonumber\\
&& \times T \Big[\ln \Big(1+\exp (-\frac{\omega^+ -(n+\frac{1}{2}) \omega }{T})\Big) \nonumber \\
&& \qquad
+\ln \Big(1+ \exp (\frac{\omega^+-(n+\frac{1}{2}) \omega }{T}) \Big) \nonumber\\
&& \qquad + \ln \Big(1+ \exp (-\frac{\omega^- -(n+\frac{1}{2}) \omega }{T}) \Big) \nonumber\\
&& \qquad +\ln \Big(1+ \exp (\frac{\omega^- -(n+\frac{1}{2}) \omega }{T})\Big) \Big]
\end{eqnarray}
where $J_n$ are n-th Bessel function of the first kind and  quasiparticle dispersion relations are given by
\begin{eqnarray}
\omega^\pm = \sqrt{4 G_s^2 \pi^2+(\sqrt{(m_0-2 G_s \sigma )^2+k_t^2+k_z^2} \pm\widetilde{\mu_I} )^2}
\end{eqnarray}
with $\widetilde{\mu_I}=\frac{\mu_I}{2}+G_v\, \rho$. 
The mean-field condensates are then numerically solved from coupled  gap equations:
\begin{eqnarray} \label{eq_Gap}
\frac{\partial \Omega}{\partial \sigma}=\frac{\partial \Omega}{\partial \pi}=\frac{\partial \Omega}{\partial \rho}=0 ,
\end{eqnarray}
Note the system under rotation is no longer homogeneous with thermodynamic quantities varying with radial coordinate $r$. For the numerical results to be presented later, we use a value $r=0.1\ {\rm GeV}^{-1}$. This is a rather modest value that ensures $\omega\, r \ll 1$ in all our calculations and that renders finite boundary effect to be negligible. The qualitative features of our findings do not depend on this particular choice and generally the rotational effect increases with larger value of $r$ (see e.g.~\cite{Ebihara:2016fwa}).

\section{\label{sec:Scalar} Rotational Suppression of Pion Condensation}

We first demonstrate the rotational suppression of scalar pairing channels. 
While such rotational suppression was previously 
demonstrated for e.g. chiral condensate or color superconductivity~\cite{Jiang:2016wvv}, it has never been examined for the mesonic condensation in isospin matter. In our case, the scalar pairing channels include the condensates of both $\sigma$ (scalar) and $\pi$ (pseudo-scalar) fields. To show this effect clearly, we will temporarily ``turn off'' the vector channel in the present section by putting $G_v$ and $\rho$ to zero in Eq.\eqref{eq_Omega}.

In the case without rotation, this system has been very well studied. At low isospin density the system is vacuum-like with only a nonzero $\sigma$ condensate which would decrease with increasing density.  At certain high enough isospin density, the $\pi$ condensate starts to form via a second-order phase transition --- a phenomenon called pion condensation~\cite{He:2005nk, He:2005sp, Kang:2013bea, Mao:2012fx, Cao:2015xja}. Here we focus on the influence of the rotation on this phenomenon.

\begin{figure}[htb!]
\includegraphics[width=220pt]{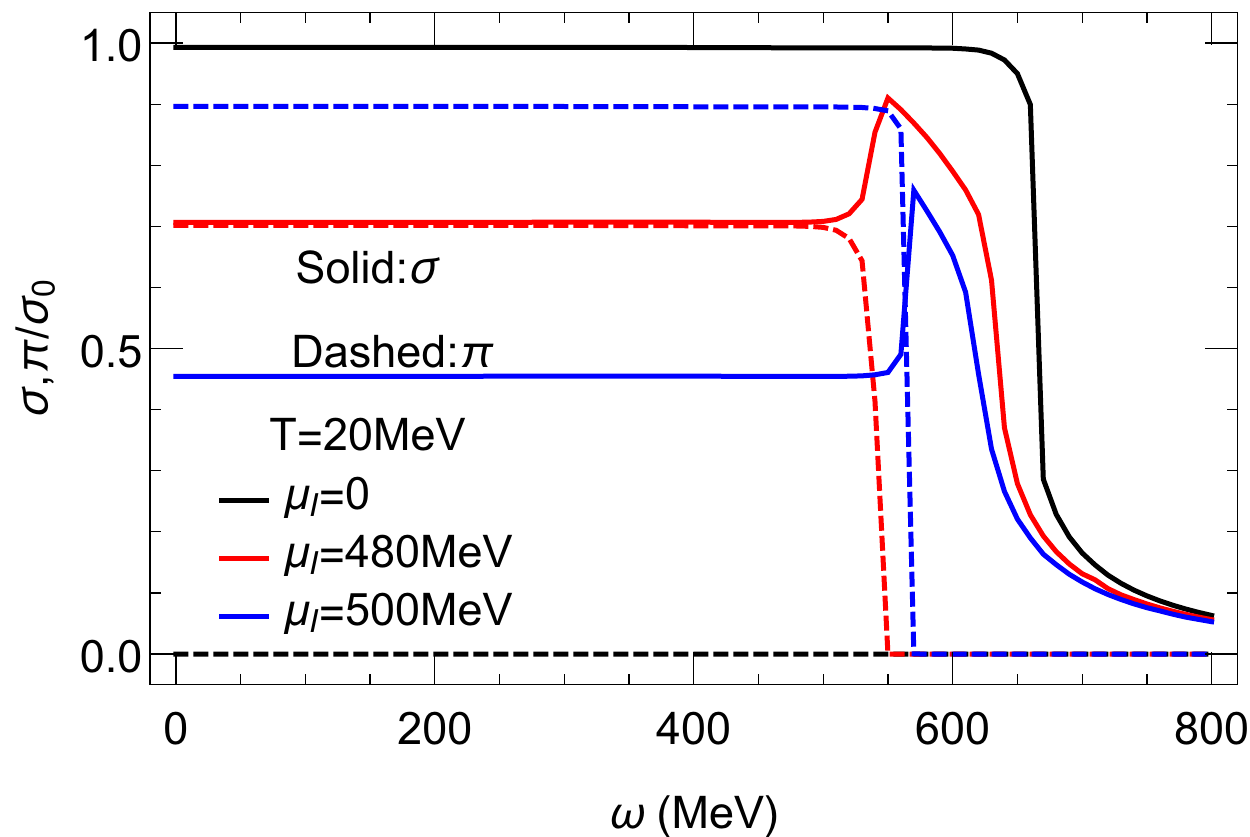}
\includegraphics[width=220pt]{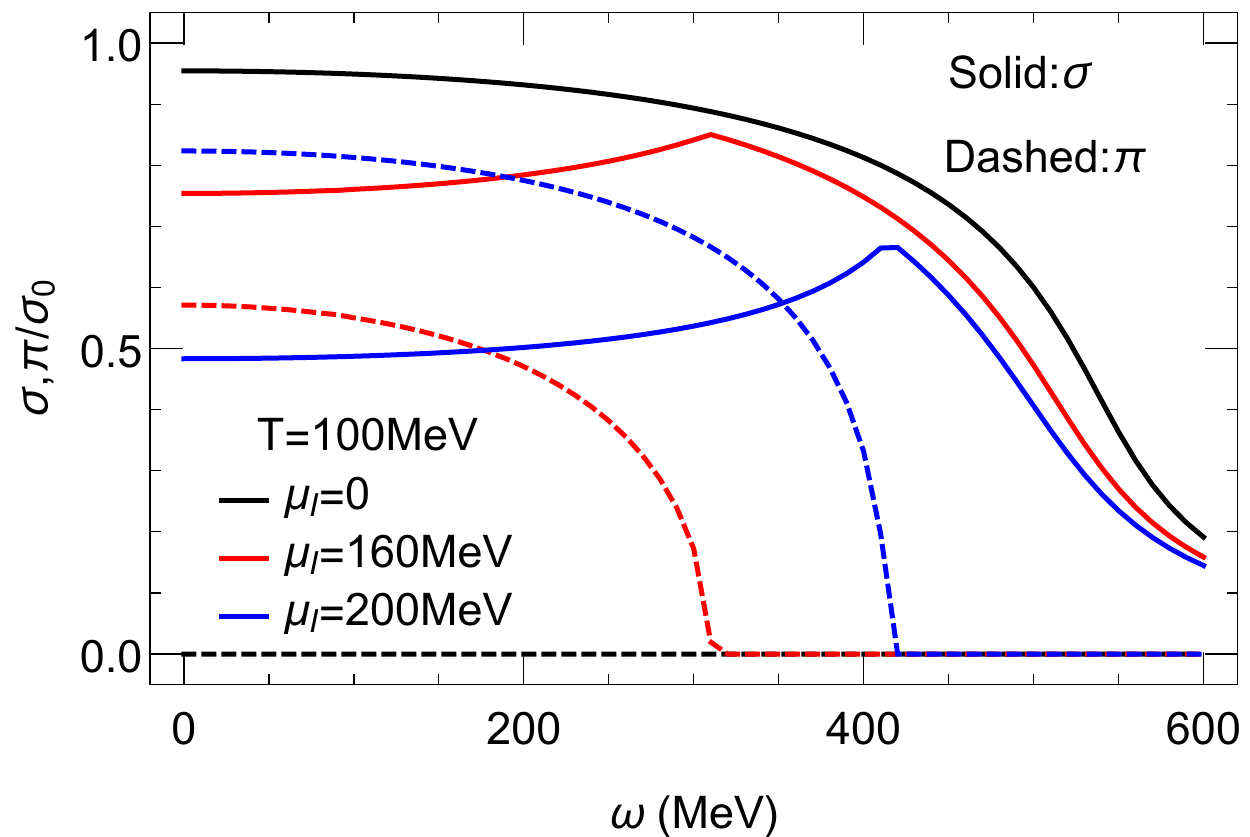}
\caption{\label{fig1} (color online) The sigma and pion condensates $\sigma$ and $\pi$ (scaled by $\sigma_0$) as a function of $\omega$ at $T=20\rm MeV$ (upper) and $T=100\rm MeV$ (lower) for several different values of $\mu_I$.}
\end{figure}

\begin{figure}[htb!]
\includegraphics[width=220pt]{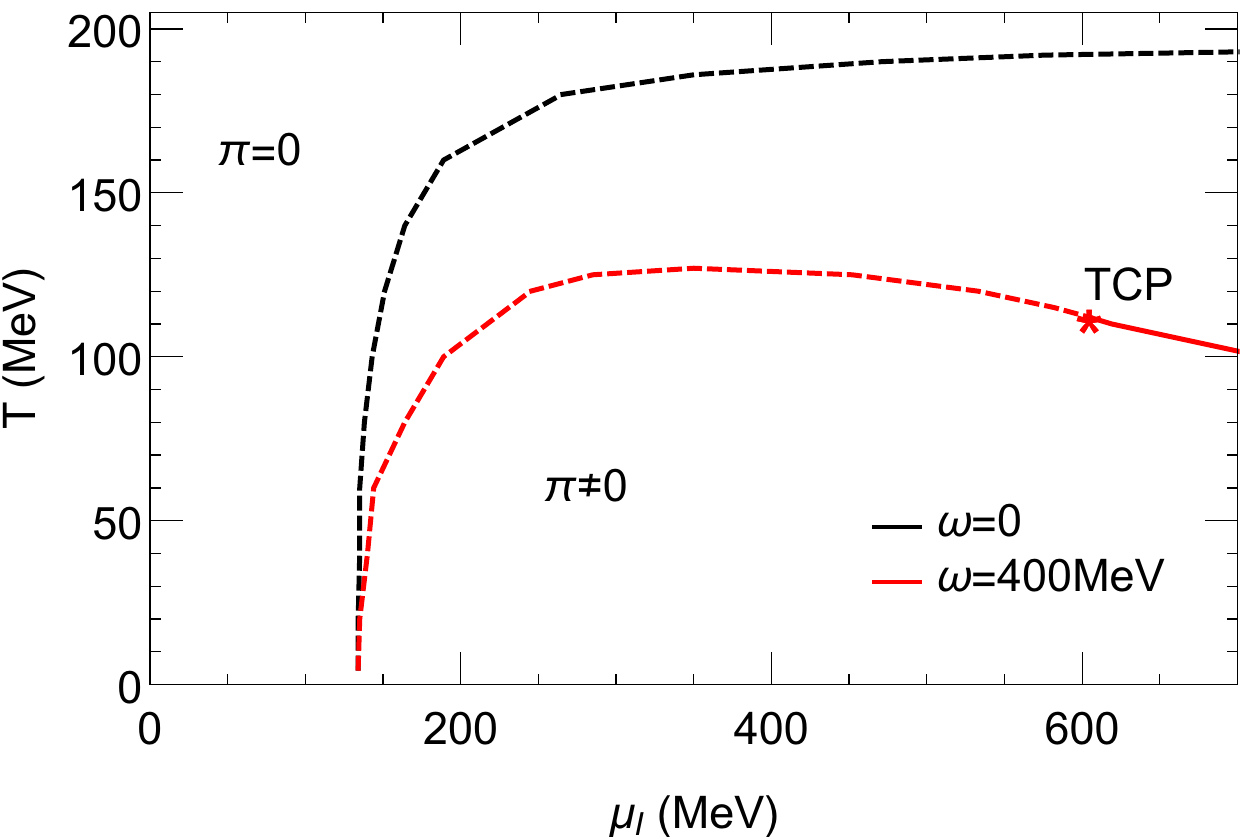}
\caption{\label{fig2} (color online) The pion condensation phase diagram on the $T-\mu_I$ plane. Dashed line stands for the second-order phase transition, while solid for the first-order. The star symbol denotes a tri-critical point (TCP).}
\end{figure}

In Fig.~\ref{fig1}, we show the sigma and pion condensates $\sigma$ and $\pi$ (scaled by $\sigma_0$) as a function of $\omega$ at $T=20\rm MeV$ (upper) and $T=100\rm MeV$ (lower) for several different values of $\mu_I$. As one can see, there is indeed a generic suppression on both the $\sigma$ and $\pi$ condensates, due the fact that the presence of global rotation always ``prefers'' states with nonzero angular momentum and thus disfavors these $J=0$ mesonic pairing channels. What's most interesting is the case at high isospin density, where the system is in a pion condensation phase with nonzero pion condensate without rotation. But with increasing rotation, this condensate eventually approaches zero via either a first-order (at low T) or second-order (at high T) transition. Thus the spontaneously broken isospin symmetry in the pion condensation phase can be restored again under rapid rotation, which is a new effect.

To see the influence of rotation on the phase structure, one can compare compare the $T-\mu_I$ phase diagram of isospin matter with and without rotation. As shown in Fig.~\ref{fig2}, the region of pion condensation phase is significantly reduced by the rotation. In particular due to the rotation, a new first-order transition line emerges at  high-$\mu_I$ side which connects to the second-order line at low-$\mu_I$ side via a new tri-critical point (TCP).

\section{\label{sec:Vector} Enhanced Rho Condensation under Rotation}

\begin{figure}[htb!]
\includegraphics[width=200pt]{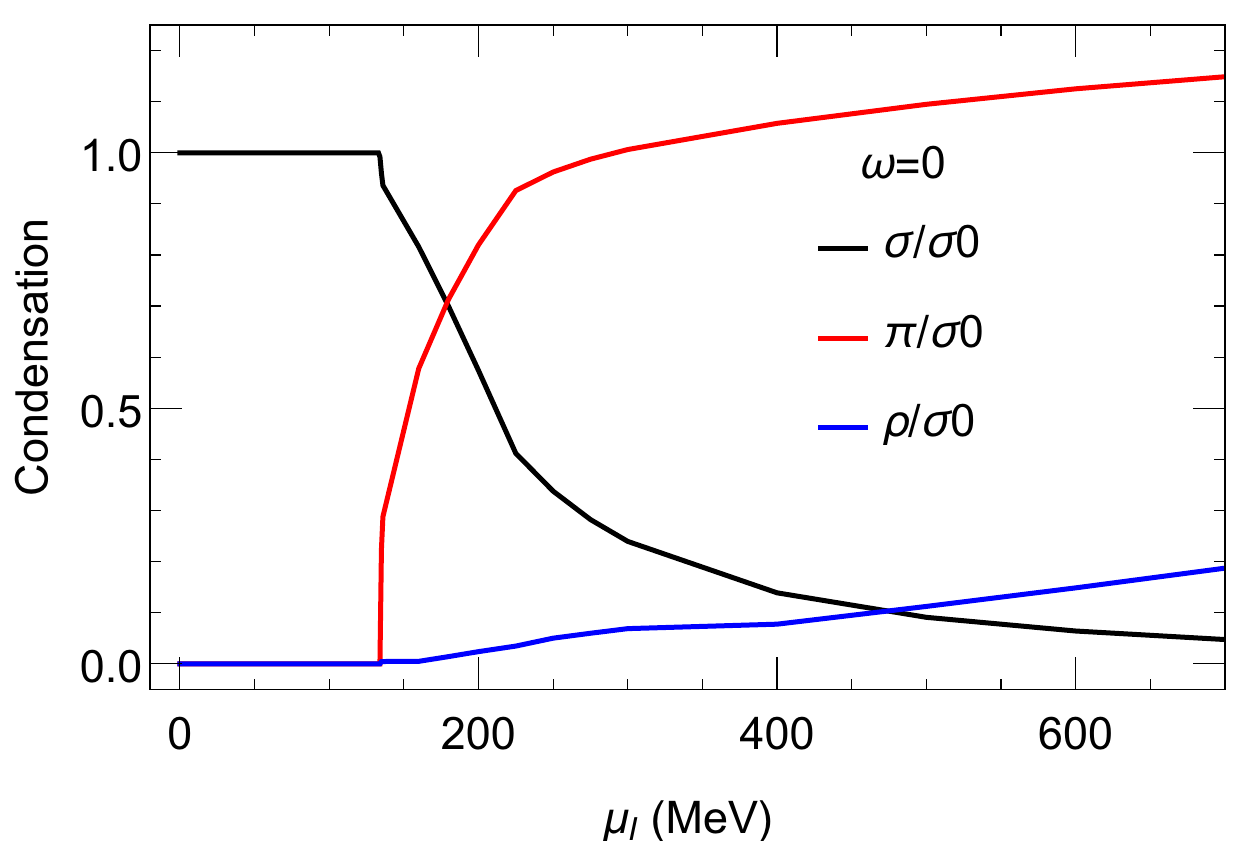}
\includegraphics[width=200pt]{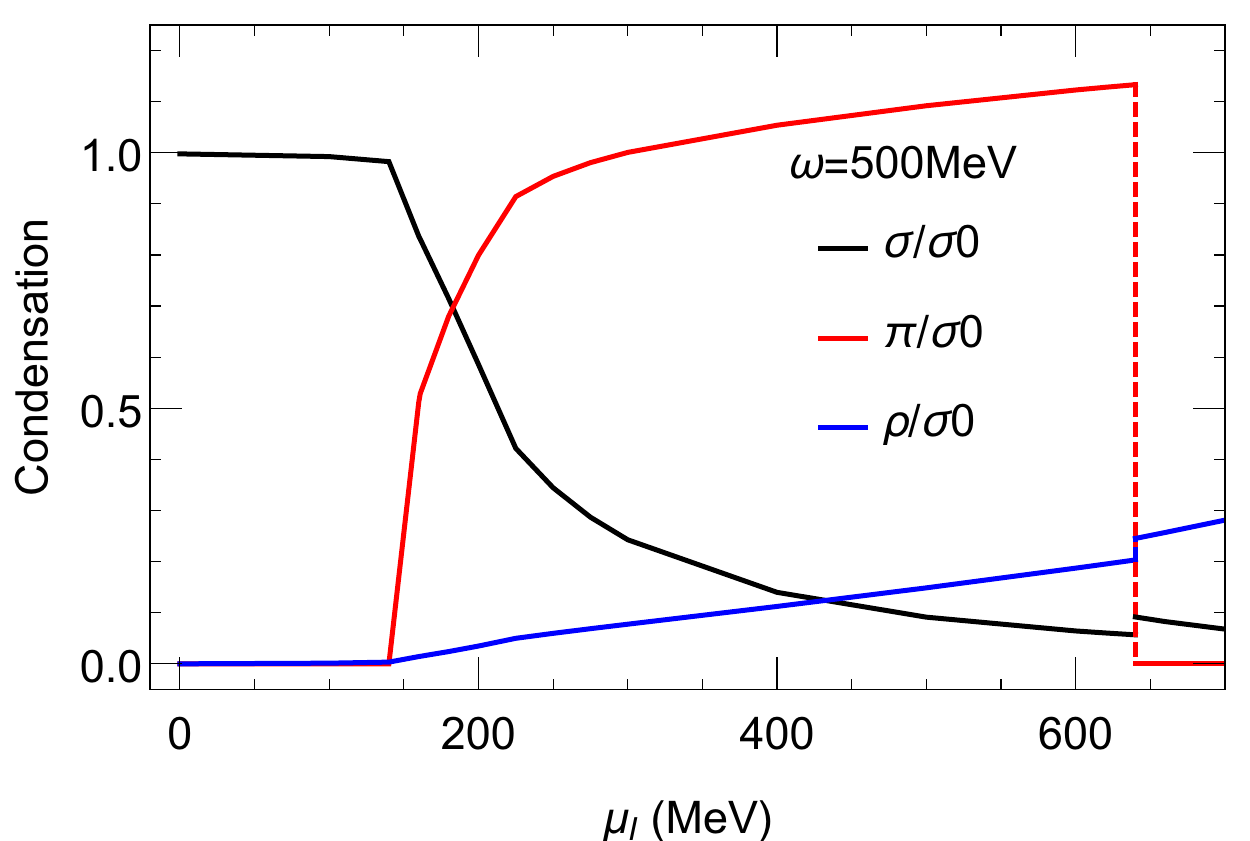}
\includegraphics[width=200pt]{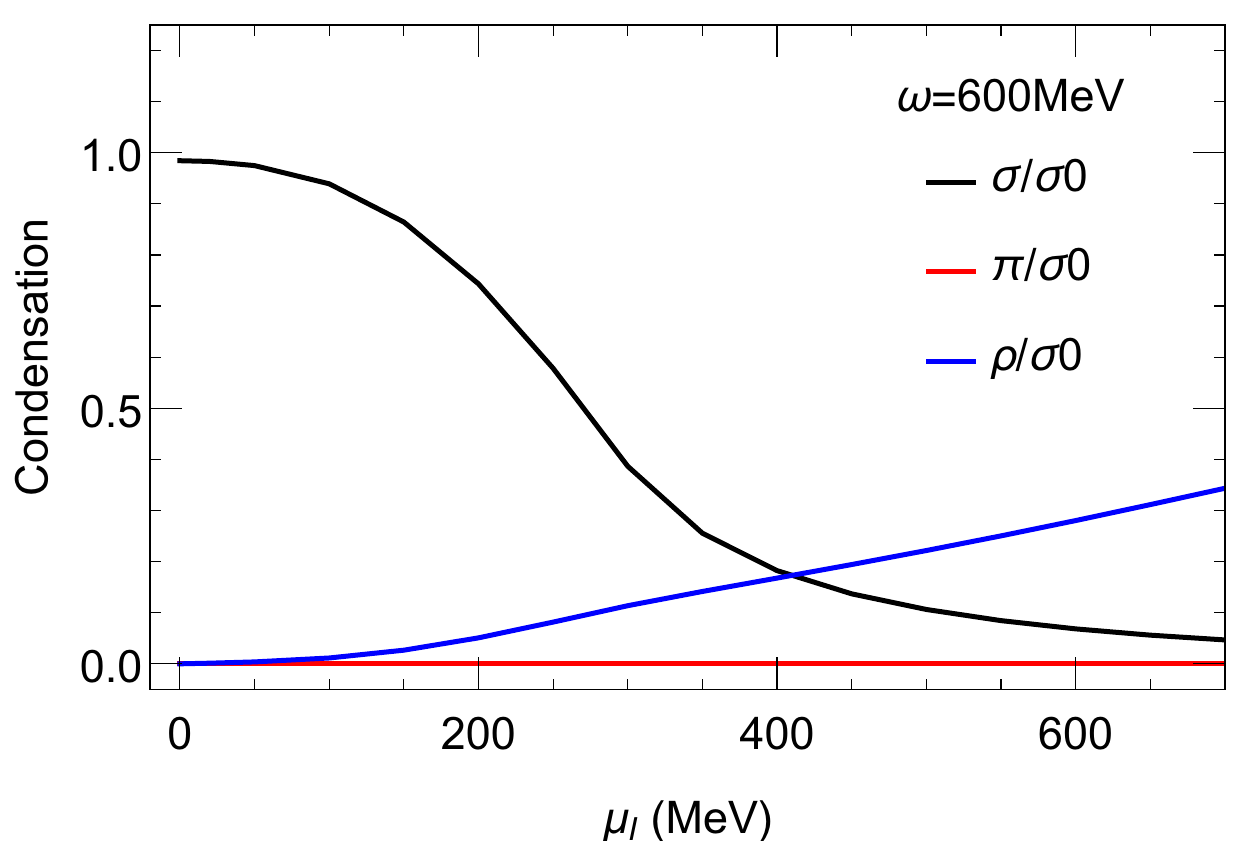}
\caption{ (color online) The sigma, pi and rho condensates $\sigma,\ \pi,\ \rho$ (scaled by the vacuum chiral condensate $\sigma_0$) as a function of isospin chemical potential, for
$\omega=0$ (upper), $\omega=500\rm MeV$ (middle) and $\omega=600\rm MeV$, respectively. The temperature is $T=10\rm MeV$.}\label{fig3}
\end{figure}

Suppression of the scalar pairing implies opportunity for enhanced pairing of states with nonzero angular momentum, such as the $\rho$ condensate. Indeed, the $\rho$ state has $J=1$ and should be favored by the presence of global rotation. While the emergence of $\rho$ condensate at high isospin density has been previously studied~\cite{Brauner:2016lkh}, the interplay between the rho condensate and rotation and the implication for phase structure of isospin matter is discussed for the first time here. To do this, we now consider the full thermodynamic potential in Eq.\eqref{eq_Omega} and consistently solve the coupled gap equations of all three possible condensates in Eq.\eqref{eq_Gap}. In Fig.\ref{fig3}, we compare the results for the sigma, pi and rho condensates $\sigma,\ \pi,\ \rho$ (scaled by the vacuum chiral condensate $\sigma_0$) as a function of isospin chemical potential, for $\omega=0$ (upper), $\omega=500\rm MeV$ (middle) and $\omega=600\rm MeV$, respectively. The temperature for this calculation is $T=10\rm MeV$.

In the case without rotation (Fig.\ref{fig3} upper panel), the chiral condensate decreases with increasing $\mu_I$  while both pion and rho condensates start to grow for $\mu_I$ greater than the critical value at about $140\rm MeV$ for a second order phase transition.  The pion condensate dominates the system at large isospin chemical potential.

In the case with strong rotation, $\omega=500\rm MeV$ (Fig.\ref{fig3} middle panel),  the situation becomes different. Both pion and rho condensates still start to grow for $\mu_I$ greater than the critical value. But at even higher isospin density, a new first-order transition occurs and the pion condensate drops to zero. In this new region, the rho condensate becomes dominant.

For even stronger rotation,  $\omega=600 \rm MeV$ (Fig.\ref{fig3} lower panel), the pion condensate disappears all together. With increasing isospin chemical potential $\mu_I$, there is a smooth crossover from a $\sigma$-dominated phase at low isospin density to a
$\rho$-dominated phase at very high isospin density.

\begin{figure}[htb!]
\includegraphics[width=220pt]{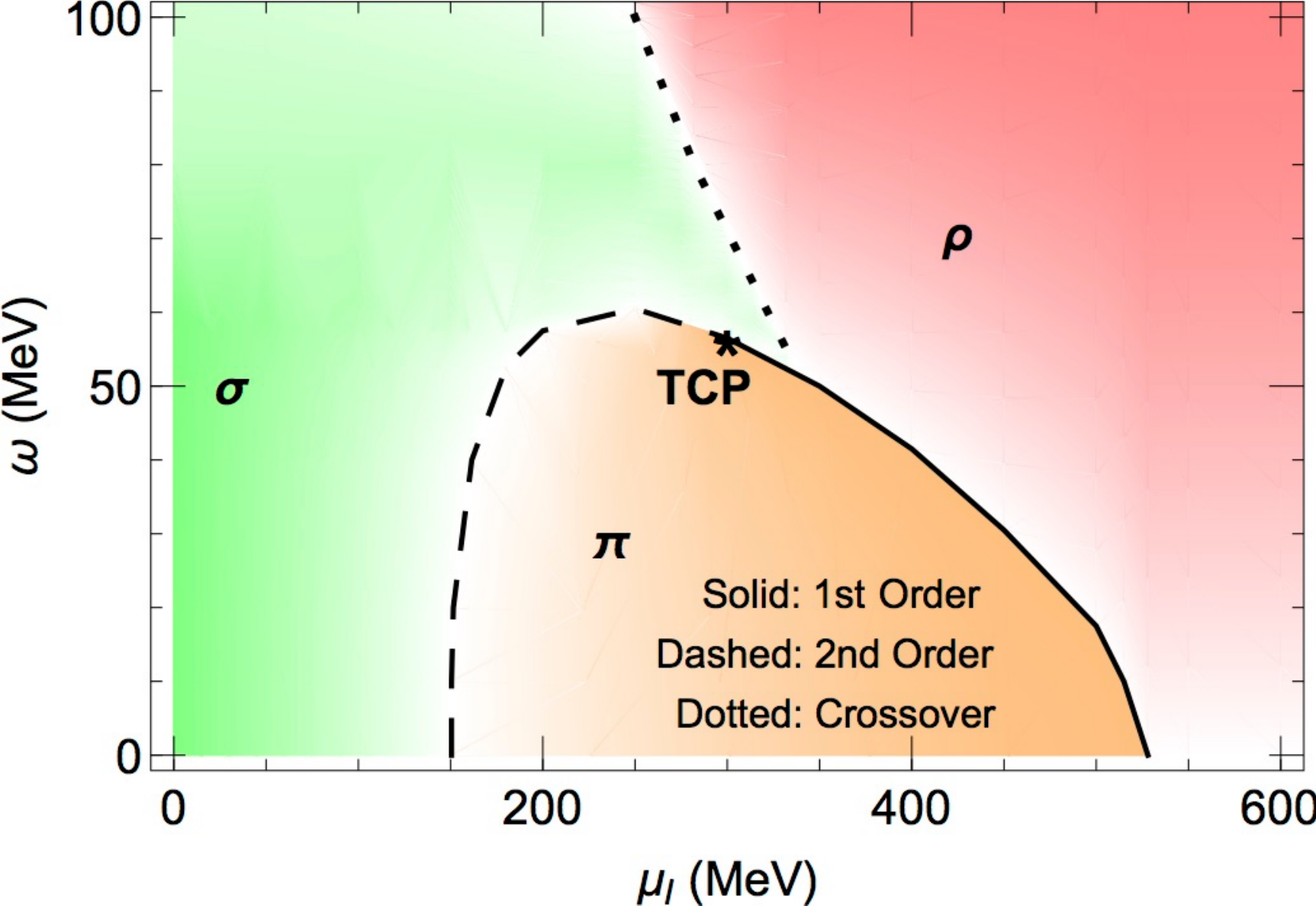}
\caption{ (color online) A new phase diagram on the $\omega-\mu_I$ plane for mesonic condensation in isospin matter under rotation.   solid line stands for first-order phase transition and dashed line  for second-order transition, while dotted line for crossover, with the star symbol denoting a tri-critical point (TCP) at  $(\mu_I^c=300{\rm MeV},\omega^c=57{\rm MeV})$. The temperature is $T=10\rm MeV$ and quark chemical potential is $\mu=250\rm MeV$.}\label{fig4} \vspace{-0.2in}
\end{figure}

These results clearly demonstrate the influence of rotation on the mesonic condensation in isospin matter and envision a new phase diagram on the $\omega-\mu_I$ plane, as shown in Fig.~\ref{fig4}. This new phase structure is characterized by three distinctive regions: a vacuum-like, sigma-dominated phase in the low isospin density and slow rotation region; a pion-condensation phase in the mid-to-high isospin density with moderate rotation; and a rho-condensation phase in the high isospin and rapid rotation region. A second-order transition line   separates the sigma-dominant and pion-dominant regions while a first-order  line   separates the pion-dominant and rho-dominant regions, with a tri-critical point connecting them. Similar phase structure has been found  at various temperatures and  chemical potentials, and is a robust feature from interplay between rotation and isospin.

\section{\label{sec:summary} Conclusion}

In this paper, we have investigated the mesonic condensation in isospin matter under rotation. Using the two-flavor NJL effective model under the presence of global rotation, we have demonstrated two important effects of the rotation on the phase structure: a rotational suppression of the scalar-channel condensates, in particular the pion condensation region; and a rotational enhancement of the rho condensation region with vector-channel condensate. A new phase diagram for isospin matter under rotation has been mapped out on the $\omega-\mu_I$ plane where three distinctive phases, corresponding to $\sigma$, $\pi$, $\rho$ dominated regions respectively, are separated by a second-order line at low isospin chemical potential and a first-order line at high rotation which are joint by a tri-critical point. While the quantitative details  may depend on model details, we expect such a three-region phase structure to be generic.

In the present study, we have not considered the  finite size effect on the phase structure~\cite{Ebihara:2016fwa}. The finite size correction may  further reduce the region of pion condensation, but the qualitative feature of the phase structure is expected to remain.  Our calculation of the thermodynamic potential with rotation is under mean field approximation, and the inclusion of   fluctuations beyond such  approximation may change the precise locations   of the tri-critical point or phase transition lines. But the influence of   rotation on the mesonic condensation, revealed in this Letter, shall remain the same.

The novel phase structure found here may be useful and relevant for understanding properties and phenomena in isospin-asymmetric nuclear matter in various physical systems.  For peripheral heavy ion collisions at relatively low beam energies, such as experiments at the RHIC Beam Energy Scan or at the future FAIR and NICA facilities~\cite{Bzdak:2019pkr},  the created matter has significant rotation as well as high isospin density due to stopping effect and the asymmetry between protons and neutrons in the initial nuclei.  Studies on  the phase structure of asymmetric dense cold nuclear matter  with rotation  at finite baryon density and low temperature will be useful for study the structure and properties of rotating compact star. Additional rich phases such as vortex lattice or inhomogeneous FFLO state~\cite{Splittorff:2000mm} in such environment are also interesting possibilities for further study. Based upon the theoretical study of the present paper, a detailed investigation of relevant real-world systems is underway and will be reported elsewhere in the future.

  \begin{acknowledgements}
  The authors thank Yin Jiang and Shuzhe Shi for useful discussions and communications. This work is in part supported by the Ministry of Science and Technology of China (MSTC) under the "973" Project No. 2015CB856904(4), by NSFC Grant No. 11735007, by NSF Grant No. PHY-1352368, PHY-1913729 and by the U.S. Department of Energy, Office of Science, Office of Nuclear Physics, within the framework of the Beam Energy Scan Theory (BEST) Topical Collaboration. HZ acknowledges partial support from the China Scholarship Council. JL is  grateful to the Institute for Advanced Study of Indiana University for partial support. 
  \end{acknowledgements}
\vspace{-0.2in}


\end{document}